\documentclass[10pt,conference,letterpaper]{IEEEtran}
\usepackage{color,graphicx,amsmath,amssymb,epsfig,subfigure,mathrsfs,cite}
\psfull

\title{Multistage Relaying Using Interference Networks}
\author{Bama Muthuramalingam, Srikrishna Bhashyam and Andrew Thangaraj\\
Department of Electrical Engineering\\
Indian Institute of Technology Madras, Chennai, India 600036 \\
\{bama,\ skrishna,\ andrew\}@iitm.ac.in}

\begin{document}
\maketitle
\begin{abstract} 
Wireless networks with multiple nodes that relay information from a
source to a destination are expected to be deployed in many
applications. Therefore, understanding their design and performance
under practical constraints is important. In this work, we propose
and study three multihopping decode and forward (MDF) protocols for
multistage half-duplex relay networks with no direct link between
the source and destination nodes. In all three protocols, we assume
no cooperation across relay nodes for encoding and
decoding. Numerical evaluation in illustrative example networks and
comparison with cheap relay cut-set bounds for half-duplex networks
show that the proposed MDF protocols approach capacity in some
ranges of channel gains. The main idea in the design of the
protocols is the use of coding in interference networks that are
created in different states or modes of a half-duplex network. Our
results suggest that multistage half-duplex relaying with practical
constraints on cooperation is comparable to point-to-point links and
full-duplex relay networks, if there are multiple non-overlapping
paths from source to destination and if suitable coding is employed
in interference network states.
\end{abstract} 

\vspace*{-2mm}
\section{Introduction} 
\label{Sec:Intro}
One of the key technologies in next generation systems for achieving high throughput and providing better coverage is \emph{relaying}. Relaying has attracted a high level of recent research interest with several papers focusing on various aspects of communicating using relays with different constraints and assumptions. In this work, we are concerned with the capacity of multistage relaying from one source to one destination through an arbitrary network of half duplex relays. 

\begin{figure}[ht]
\centering
\mbox{
\subfigure
{\resizebox{1.4in}{!}{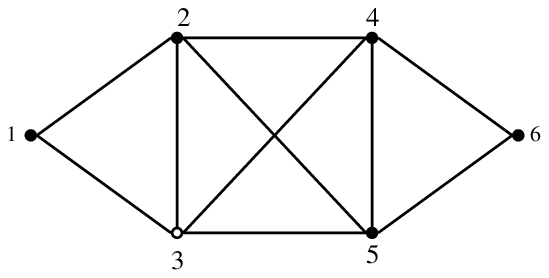} 
} 
\subfigure
{\resizebox{1.8in}{!}{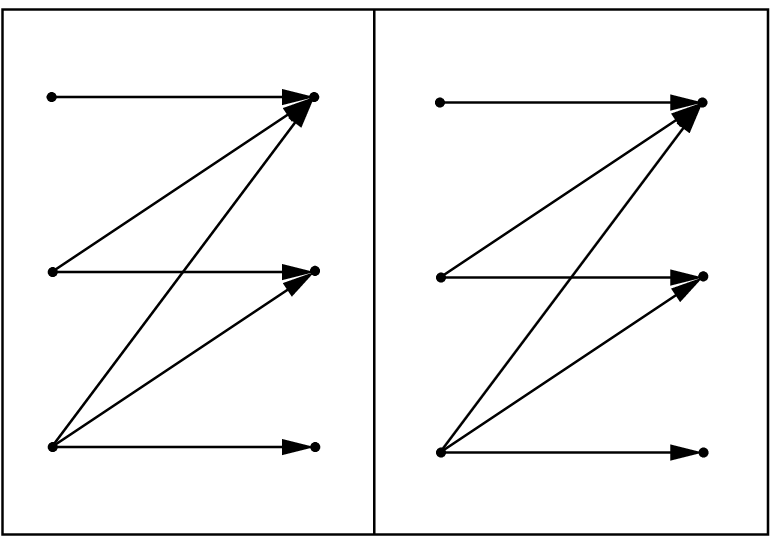}
}
}
\caption{Two stage relay network and interference states}
\label{Fig:LNET}
\end{figure}

An example network that we consider in detail for ease of explanation and clarity is the two stage relay network shown in Fig. \ref{Fig:LNET}. In this 6-node network, the source node $S=1$ intends to communicate with the sink node $D=6$ through 4 relay nodes $\{R_1=2,R_2=3,R_3=4,R_4=5\}$ connected as shown. 
The channel gains ($\alpha$, $\beta$, $\gamma$) are shown next to the corresponding edges. For simplicity, some of the gains are assumed to be identical. For a multistage half-duplex relay network such as the one in Fig. \ref{Fig:LNET}, we study coding methods and protocols needed to achieve the best possible rate from source to destination for different ranges of the channel gains. 

There are two different aspects to multistage relaying when the relays are connected in an arbitrary fashion: (1) scheduling transmissions by nodes, and (2) coding methods employed by nodes during transmissions. One strategy for scheduling is to avoid interference altogether. However, the maximum data rate under Interference Avoidance (IA) is limited, because the source is transmitting only for a fraction of the total time. To improve upon IA, more states of the network with the source node in transmit mode need to be considered. 
The scheduling task is to determine those states that are crucial for obtaining higher rates. When multiple nodes transmit, interference network states are created in the network based on the connectivity. Two important interference network states are shown in Fig. \ref{Fig:LNET} for the network of Fig. \ref{Fig:LNET}. In one state, $S$, $R_1$, and $R_3$ are transmitters and, in the other state, $S$, $R_2$, and $R_4$ are transmitters. In both the states shown in Fig. \ref{Fig:LNET}, the source node is a transmitter and the destination node is a receiver. This property improves the flow of information, and is useful for improving the transmission rate from the source.

In each interference network state, three different coding strategies of increasing complexity are considered for transmitters - Common broadcast (CB), Superposition coding (SC) and Dirty paper coding (DPC) for the source node alone. The receiving nodes in the interference network employ multiple access (MAC) receivers that work by successive interference cancellation. For different combinations of coding strategies, suitable rate regions are determined for each state (or interference network). The overall rate achievable from the source to the destination is computed using an optimization over the time-sharing of the rate regions for each state, subject to additional flow constraints that ensure compatibility of the rate vectors used for individual states. 

To place our work better, we review a sample of the relevant prior literature. The relay channel is a classic setting, introduced in \cite{VandeRelay}, and studied extensively \cite{CoverRelay,Laneman03,Kramer05}. One result of particular interest is the cut-set bound for half-duplex relay networks operating by time-sharing over a finite number of states \cite{BoundAmir}. This ``cheap relay'' bound has been used by several authors as an outer bound for achievable rates.


Recently, the half-duplex diamond network with two relays has been studied in \cite{XueDiamond, AmirDiamond, ChangDiamond, RankovDiamond}. The \emph{multi-hopping decode and forward} (MDF) protocol, proposed in \cite{XueDiamond} and extended in \cite{AmirDiamond}, achieves rates close to the cheap relay cut-set bound. Wang \emph{et al} \cite{ChangDiamond} consider a modified diamond network with an additional link between the relays and propose a coding strategy using Dirty Paper Coding (DPC), which is shown to approach the cut-set bound. More protocols for general half-duplex wireless relay networks have been studied in \cite{DebdeepGeneral}, \cite{FettweisGeneral}.


In relation to the above, in our work, we propose and study multi-hopping decode and forward (MDF) protocols for a general relay network with half-duplex nodes in the following setting: (1) \emph{No cooperation} is assumed for encoding and decoding (except in one protocol for the source node alone), (2) Achievable rates are compared against the cheap relay cut-set bound at \emph{finite SNRs}, (3) The protocols and methods apply for a \emph{general topology} of relays. The results are illustrated by evaluation on two different networks, where we show that the cut-set bound is approached for some values of channel gains. 


\vspace*{-2mm}
\section{Model}
\label{Sec:Model}
We represent a wireless network with $m$ nodes as an undirected graph $G=(V,E)$, where the vertex set $V =\{ 1,2,\ldots, m\}$ represents the wireless nodes. An edge $(i,j)\in E$ indicates that Node $i$ and Node $j$ are connected by an additive white Gaussian noise (AWGN) channel with constant gain denoted as $h_{ij}$.

Each node is subject to an average power constraint $P$ and has a noise variance $\sigma^2$. In addition, a half-duplex constraint is imposed on the nodes so that they can either transmit, receive, or be idle at any given time. Therefore, in this work, an $m$-node half-duplex wireless network can be in $ M \leq {\mathscr M} = 3^m$ states. These states are denoted $ S_1$, $ S_2$, $\cdots$, $ S_{ M}$. In such a network, we are interested in maximizing the communication rate $R$ from an arbitrary source $S\in  V$ to an arbitrary sink $D \in  V$. The nodes in $ V\setminus \{S,D\}$ act as relays in this communication. Information flow from source to destination happens by a time-sharing of the states $ S_k$, $1\leq k\leq  M$, and may reach the destination in multiple hops depending on the connectivity of the graph. Hence, the specific problem considered in this work can be termed {\it multihop, half-duplex relaying} in an arbitrary wireless network.

The total transmission time is normalized to one time unit, and state $ S_k$ is active for a $\lambda_k$ fraction of the time ($\lambda_k$ could be zero) with $\sum_{k=1}^{ M}\lambda_k = 1$. As in \cite{XueDiamond,AmirDiamond}, we assume that the state sequence and the time-sharing parameters are known to all nodes before transmission. Let $I_k =\{i \in  V: \text{Node} \;i\;\text{is a transmitter in State} \;  S_k \} $ be the set of active transmitters in State $ S_k$, and let $J_k =\{i \in  V: \text{Node} \;i\; \text{is a receiver in State} \;  S_k \}$ be the set of active receivers in State $ S_k$. When state $ S_k$ is active, simultaneous transmissions from nodes in $I_k$ can interfere at one or more of the receivers in $J_k$ depending on the connectivity of the nodes in $I_k$ and $J_k$. Thus, each state $ S_k = (I_k, J_k)$ is an {\em interference network} \cite{carleial1978interference} or {\em hyperedge} with $I_k$ and $J_k$ as the two disjoint vertex sets. We use the terms interference network, hyperedge and state interchangeably. The choice of a specific coding and decoding strategy for each state $S_k=(I_k,J_k)$ determines possible operating rate vectors in an achievable rate region for that state. Since the capacity region and optimal coding
scheme are not known for general interference networks, we consider three suboptimal strategies for each state based on different broadcast and interference processing techniques. In all these strategies, we impose the constraint that the receivers $J_k$ cannot cooperate in decoding. Similarly, the nodes in $I_k$ are assumed to encode their messages independently; however, in one scheme, the source is assumed to know the messages transmitted by the relays.

\vspace*{-2mm}
\section{Cut-set Bound}
\label{Sec:UB}
A cut-set upper bound for half-duplex relay networks operating by
time-sharing over a finite number of states has been derived in
\cite{BoundAmir}. This bound is presented here,
briefly.

Let $X^{(i)}$ and $Y^{(i)}$ be the transmitted and received variables
at node $i$ when it is in transmit and receive states, respectively.
The maximum achievable information rate $R$ between source $S$ and
destination $D$ in a half-duplex network is bounded as
\begin{equation}
  R \leq \;\sup_{\lambda_k} \; \min_{\Omega} \;\sum_{k=1}^{{\mathscr M}} \lambda_k I(X_{(k)}^\Omega ; Y_{(k)}^{\Omega ^c}|X_{(k)}^{\Omega ^c}),
\label{eq:HDUB}
\end{equation}
for some joint distributions $\{ p(x^{(1)}, x^{(2)}, \cdots, x^{(m)} |
k)\},$ $1 \le k \le {\mathscr M}$, where the supremum is over all
$\lambda_k \geq 0$ such that $\sum_{k=1}^{\mathscr M} \lambda_k = 1$,
the minimization is over all $\Omega$ such that $S \in \Omega$, $D \in
\Omega^c$, $X_{(k)} ^\Omega = \{X^{(i)}: i\in \Omega \cap I_k\}$,
$Y_{(k)}^{\Omega ^c} =\{Y^{(i)}: i\in \Omega ^c \cap J_k\}$, and
$X_{(k)}^{\Omega ^c} =\{X^{(i)}: i\in \Omega ^c \cap I_k\}$. The above
upper bound can be computed by solving a linear program
\cite{AmirDiamond}. The mutual information $I(X_{(k)}^\Omega ;
Y_{(k)}^{\Omega ^c}|X_{(k)}^{\Omega ^c})$ is computed exactly using known sum rate
capacity results \cite{CoverBook} when the choice of $\Omega$ and $k$
results in multiple access or broadcast channels. When the sum rate
capacity is not known exactly (e.g. for interference channels), the
MIMO sum capacity is used as an upper bound.

\vspace*{-1mm}
\section{Multihop Half-duplex Relaying Strategies}
\label{Sec:AchRate}
In this section, we present the three MDF strategies that we propose
and study in the context of a general relay network with half-duplex
nodes. In all these strategies, the network operates by time-sharing
between the states, where each state is an interference network in
general. The strategies differ in the encoding scheme in each
state. The decoder at each receiver employs successive interference
cancellation (SIC).  

\subsection{Common Broadcast (CB) Scheme}
In state $S_k=(I_k,J_k)$, each transmitter $ i \in I_k$ sends a common
message at rate $R^k_i$ to the set of all its receivers
$\Gamma_{-}^i$. Each receiver $j \in J_k$ must decode the messages
from the set $\Gamma_{+}^j$ of all the transmitters connected to
$j$. The decoding constraints at each receiver for achievability are
the constraints for the multiple access channel corresponding to the
SIC receiver. Therefore, the achievable rate region for each state
$S_k$ is defined by the following set of constraints:
\begin{equation}
\sum_{i \in A} R_i^k \le \frac{1}{2}\log \left(1 + \frac{\sum_{i \in A} h^2_{ij}P}{\sigma^2} \right),
\label{CBconstraint} 
\end{equation}
for all $A \subseteq \Gamma_{+}^j$ and for all $j \in J_k$. When
each transmitter is connected to all receivers, i.e., $\Gamma_{-}^i
= J_k$ for each $i \in I_k$, then the above region is the same as
the compound multiple access rate region in \cite{HanMAC}.

\subsection{Superposition Coding (SC) Scheme}
In this scheme, in state $S_k$, each transmitter $i \in I_k$ sends
independent messages to each of its receivers in $\Gamma_{-}^i$ using
superposition coding. Let the codeword transmitted to receiver $j$
from transmitter $i$ be ${\mathbf x}_{ij}$. Let the power used for
this codeword be $P_j = \alpha_{ij} P$ and  $R_{ij}^k$ be the rate.  Therefore, the transmitter $i$
transmits a superposition of codewords given by ${\mathbf x}_{i} =
\sum_{j \in \Gamma_{-}^i}{\mathbf x}_{ij}$. The received word at
receiver $j$ is 
\[
{\mathbf y}_j = \sum_{i\in \Gamma_{+}^j} h_{ij}\sum_{l \in \Gamma_{-}^i} {\mathbf x}_{il} + {\mathbf w}_j.
\]
For simplicity of notation, we assume that the $d_-^i$ receivers in
$\Gamma_{-}^i$ are arranged in descending order of channel magnitude
from transmitter $i$, and $\Gamma_{-}^i[p:q]$ denotes the set of elements of
$\Gamma_{-}^i$ starting from the $p^{th}$ element to the $q^{th}$
element. Each receiver $j$ decodes the codewords intended for itself
and all other {\em weaker} receivers from each transmitter. Let
receiver $j$ be the $l_i^{th}$ receiver in $\Gamma_-^i$. The codewords
of the weaker receivers $\Gamma_-^{i}[l_i+1:d_-^i]$ are canceled in
the SIC receiver. Therefore, only the codewords to the stronger
receivers $\Gamma_-^{i}[1:l_i -1]$ will interfere. The received
word can be written as
\begin{eqnarray*}
{\mathbf y}_j &= &\underbrace{\sum_{i\in \Gamma_+^j} \sum_{l \in \Gamma_-^i[1:l_i-1]} h_{ij}{\mathbf x}_{il}}_{\text{interference codewords}} \\ & + &\underbrace{\sum_{i\in \Gamma_+^j} h_{ij} {\mathbf x}_{ij} + \sum_{i\in \Gamma_+^j} \sum_{l \in \Gamma_-^i[l_i+1:d_-^i]} h_{ij} {\mathbf x}_{il}}_{\text{decoded codewords}}+  {\mathbf w}_j.
\end{eqnarray*}
Therefore, the achievable rate region for each state $S_k$
is defined by the following set of constraints:
\begin{eqnarray}
R_{ij}^k &\le& \frac{1}{2}\log \left(1 + \frac{h^2_{ij} \alpha_{ij} P }{\sigma^2 + \displaystyle{\sum_{l \in \Gamma_-^i[1:l_i -1]}} h_{ij}^2 \alpha_{il} P} \right),
\label{SCconstraint1} \\
\sum_{j \in \Gamma_-^i} \alpha_{ij}&\le&1,\;\;\;\; \forall i \in I_k,
\label{SCconstraint2} \\
\sum_{(p,q) \in A} R_{pq}^k &\le& \frac{1}{2}\log \left(1 + \frac{\displaystyle{\sum_{(p,q) \in A} h^2_{pj} \alpha_{pq} P}}{\sigma^2 + \displaystyle{\sum_{i \in \Gamma_+^j} \sum_{l \in \Gamma_-^i[1:l_i - 1]} h_{il}^2 \alpha_{il} P}} \right),
\label{SCconstraint3} 
\end{eqnarray}
$\forall A \subseteq Q_j = \{(p,q): p \in \Gamma_{+}^j, q \in \Gamma_{-}^p[l_i:d_-^i]\}$ and $\forall j \in J_k$.

Using superposition coding allows each transmitter to send messages to
a subset of its receivers. This {\em receiver selection} ability 
allows better spatial reuse.

\subsection{Dirty Paper Coding (DPC) - CB Scheme}
In the DPC-CB scheme, the source is assumed to know the messages
transmitted by all the relays since all messages originate from the
source. Therefore, when $S \in I_k$, Dirty Paper Coding (DPC) is used
by the source to cancel interference to its receiver caused by
simultaneous transmissions from relay nodes. Other transmitters in
$I_k$ transmit common messages similar to the CB scheme. The receiver
$r$ to which the source is sending its DPC-coded message at rate
$R_s^k$ is not affected by interference from other relays and will
decode only this message. The other receivers must decode all the
messages from all the transmitters (except the source) that are
connected to it. For example, in the state $S_1$ shown in
Fig. \ref{Fig:LNET}, $S$ transmits a DPC-coded message to $R_2$
using its prior knowledge of the messages transmitted by $R_1$ and
$R_3$ (and the corresponding channel gains). Receiver $R_4$ decodes
the common messages transmitted by $R_1$ and $R_3$, and receiver $D$
decodes the common message transmitted by $R_3$. For the above DPC-CB
scheme, the achievable rate region for state $S_k$ is given by the
following constraints:
\begin{eqnarray}
R_s^k &\le& \frac{1}{2}\log \left(1 + \frac{h^2_{sr}P}{\sigma^2} \right),
\label{DPCCBconstraint1} \\
\sum_{i \in A} R_i^k &\le& \frac{1}{2}\log \left(1 + \frac{\sum_{i \in A} h^2_{ij}P}{\sigma^2} \right), \binom{\forall A \subseteq \Gamma_{+}^j}{\forall j \in J_k \setminus r}.
\label{DPCCBconstraint2} 
\end{eqnarray}

\subsection{Optimization Model}
Now, we present the optimization problem to be solved to compute the
achievable rate from source $S$ to destination $D$ in the multistage
relay network. The optimization model from \cite{Lunadhoc} is adapted
to incorporate the appropriate rate region constraints for the MDF
schemes proposed earlier.

Let $x_{ij}^k$ denote the information flow rate from node $i$ to node
$j$ in state $S_k$ towards the sink. Let $x_i^k$ denote the total information flow out of node $i$ in state $S_k$. The optimization problem can be stated as:
\begin{equation}
\max_{\{ x_{ij}^k\}, \{ \lambda_k\}} R,\;\text{subject to:}
\label{opt-prob}
\end{equation}
\begin{itemize}
\item Flow constraints: For all $i \in V$, we have
\[
\sum_{\{k: i \in I_k\}} \sum_{j \in \Gamma_-^i} x_{ij}^k - \sum_{\{k: i \in J_k\}} \sum_{j \in \Gamma_+^i} x_{ji}^k = \left\{
\begin{array}{ll}
R & \mbox{if~} i = S \\
-R & \mbox{if~} i = D \\
0 & \mbox{else} \\
\end{array}
\right..
\]
\item Scheduling constraints: $\sum_k \lambda_k \le 1$ and $\lambda_k \ge 0$ $\forall k$.
\item Rate region constraints: The achievable rate region constraints for each state depend on the encoding and decoding scheme used. The rate constraints for each of the three proposed schemes for each state $S_k$ are as follows:
\begin{enumerate}
\item CB scheme:
\begin{eqnarray}
\sum_{j \in \Gamma_-^i} x_{ij}^k &\le& x_i^k, \mbox{~~} \forall i \in I_k,\\
\sum_{i \in A} x_i^k &\le& \lambda_k (\mbox{RHS of (\ref{CBconstraint})}), \binom{\forall A \subseteq \Gamma_{+}^j}{\forall j \in J_k},
\end{eqnarray}
where RHS of (\ref{CBconstraint}) is the right hand side of (\ref{CBconstraint}). 
\item SC scheme: Equation (\ref{SCconstraint2}), and:
\begin{eqnarray}
x_{ij}^k &\le& \lambda_k (\mbox{RHS of (\ref{SCconstraint1})}), \forall i \in I_k,\\
\sum_{(p,q) \in A} x_{pq}^k &\le& \lambda_k (\mbox{RHS of (\ref{SCconstraint3})}),
\end{eqnarray}
for all $A \subseteq Q_j$ and for all $j \in J_k$.
\item DPC-CB scheme:
\begin{eqnarray}
\sum_{j \in \Gamma_-^i} x_{ij}^k &\le& x_i^k, \mbox{~~} \forall i \in I_k,\\
x_s^k &\le& \lambda_k (\mbox{RHS of (\ref{DPCCBconstraint1})}),\\
\sum_{i \in A} x_i^k &\le& \lambda_k (\mbox{RHS of (\ref{DPCCBconstraint2})}),
\end{eqnarray}
for all $A \subseteq \Gamma_{+}^j$ and for all $j \in J_k \setminus r$.  
\end{enumerate}  
\end{itemize}

For the CB and DPC-CB schemes, the above optimization problem is a
linear program. However, for the SC scheme, it is not a linear program
since the power sharing variables $\alpha_{ij}$'s are also
optimized. Therefore, the numerical solutions for the SC scheme are
computed using the constrained optimization function {\em fmincon} in
MATLAB.

\vspace*{-2mm}
\section{Numerical Results}
\label{Sec:NumEval}

We evaluate and compare the rate achieved by the MDF schemes: (1) CB,
(2) SC, and (3) DPC-CB for two different network topologies and
channel realizations. The cheap relay cut-set upper bound for
half-duplex relay networks and the rate achieved by the IA scheme are
also evaluated. The rate achieved by each scheme is
obtained by solving the optimization problem in (\ref{opt-prob}) with
appropriate rate region constraints.

Since the diamond network has been studied in detail in
\cite{AmirDiamond,ChangDiamond}, we skip details and simply
mention that the proposed MDF protocols recover similar results for the diamond network.

\vspace*{-1mm}
\subsection{Two stage relay network}
\vspace*{-6mm}
\begin{figure}[htb]
\centering
\includegraphics[width = 3.2in]{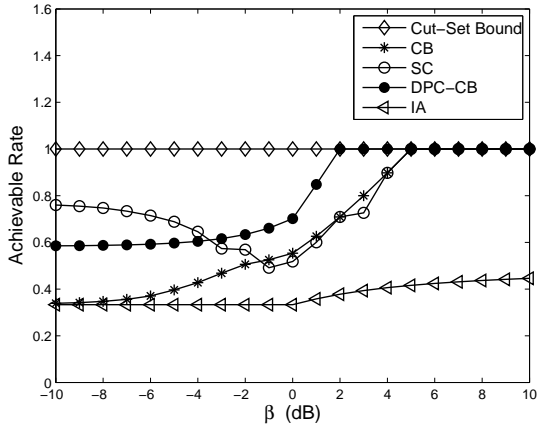} 
\caption{$\alpha = 1, \gamma = 1$, vary $\beta$.}
\label{Fig:LNETBeta} 
\end{figure} 
We first consider the two-stage relay network shown in
Fig. \ref{Fig:LNET}. For evaluating the cut-set bound, all the
$2^2\cdot3^4=324$ states were considered. The states that avoid
interference (called IA states) are the states with a single
transmitting node. For the proposed MDF protocols, interference
network states with two transmitters ($\binom{5}{2}=10$ states) and
some states with three transmitters (5 out of $\binom{5}{3}=10$ states) are
used along with the IA states. Two of the states with three
transmitters are shown in Fig. \ref{Fig:LNET} for
illustration. In Fig. \ref{Fig:LNETBeta}, the cut-set bound, determined by the source
cut, is at 1 for all $\beta$. For large $\beta$, the states used are
$S_3=(\{S,R_2,R_3\},\{R_1,R_4,D\})$ and
$S_4=(\{S,R_1,R_4\},\{R_2,R_3,D\})$. The receivers in both these
states see strong interference, which can be canceled at the
receiver. For instance, in state $S_3$, the receiver $R_1$ can decode
the source's message in the presence of strong interference from $R_2$
and $R_3$. Because of this, all three MDF schemes achieve capacity of
1 by equal time-sharing of states $S_3$ and $S_4$. For small $\beta$,
common broadcast at the relays is limited by a weak receiver with
close-to-zero capacity. Superposition coding, which enables different
rates to receivers, proves to be better at low values of $\beta$. For
SC, states $S_1$ and $S_2$ (shown in fig. \ref{Fig:LNET}) are
chosen, and the rate is limited by the interference at relays $R_1$
and $R_2$. DPC is marginally weaker, since the relays continue to do
common broadcast when the source does DPC. However, when $\beta=1$ (0
dB), DPC is better as SC becomes identical to CB for identical channel
gains.
\begin{figure}[htb]
\centering
\includegraphics[width = 3.2in]{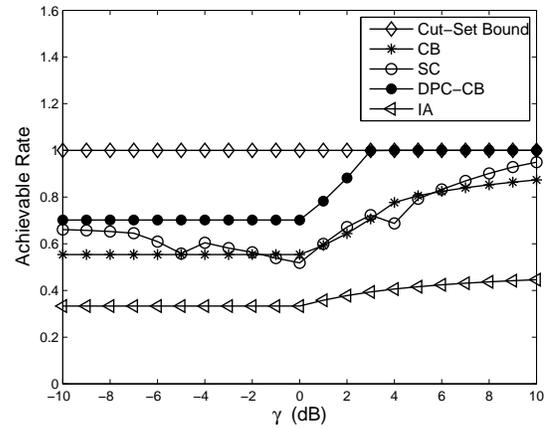} 
\caption{$\alpha = 1,\beta = 1$, vary $\gamma$.}
\label{Fig:LNETGamma} 
\end{figure} 

\vspace*{-2mm}
In Fig. \ref{Fig:LNETGamma}, the cut-set bound, determined by the
source cut, is at 1 for all $\gamma$. For $\gamma > 1$, the DPC-CB
scheme achieves the cut-set bound for lower $\gamma$ than SC and CB
schemes. DPC achieves capacity by time-sharing the states $S_1$ and
$S_2$. The interference at relays $R_1$ and $R_2$ are canceled using
DPC, while the interference at $R_3$ and $R_4$ is overcome because the
gains of the $R_2 \rightarrow R_3$ and $R_1 \rightarrow R_4$ links
increase with $\gamma$. The same states are used for the SC scheme as
well. However, interference at $R_1$ and $R_2$ are overcome only for
larger $\gamma$. For very large $\gamma$, CB scheme also approaches
the cut-set bound by time-sharing between the states
$(\{S,R_4\},\{R_1,D\})$ and $(\{R_1\},\{R_4\})$. For small $\gamma$,
we see that DPC-CB achieves a rate of 0.7, while the SC and CB
achieve rates of $0.67$ and $0.55$ respectively. For both DPC-CB and
CB schemes, states $S_3$ and $S_4$ are chosen. While the interference
at $R_3$ and $R_4$ limits the DPC-CB scheme, the CB scheme is limited
by the interference at relays $R_1$ and $R_2$.


\subsection{Rectangular grid network}
\begin{figure}[ht]
\centering
\resizebox{1.5in}{!}{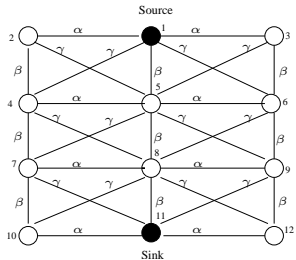}
\caption{$4 \times 3 $ Grid Network.}
\label{Fig:Grid}
\end{figure}

Consider the $4\times3$ rectangular grid network shown in
Fig. \ref{Fig:Grid}. Since the number of possible states is prohibitively
large, we first select three non-overlapping paths from the source
node $S=2$ to the destination node $D=11$. We know from the two-stage
relay network example that multiple flow paths used appropriately with
interference processing can be effective. The paths chosen are
$S\rightarrow4\rightarrow7\rightarrow D$,
$S\rightarrow5\rightarrow8\rightarrow D$ and
$S\rightarrow6\rightarrow9\rightarrow D$. Using the nodes on these
paths, the three states chosen for scheduling are $(\{S, 6 , 8\},\{4,
9, D\})$, $(\{S, 4 , 9\},\{5, 7, D\})$, and $(\{S, 5 , 7\},\{6, 8,
D\})$. Note that the source node is a transmitter and the destination
node is a receiver in all three chosen states. Also, the other two
transmitters are chosen to be at different distances from the
source. With this choice of states, we have a two-stage relay network
with six relay nodes $\{4,5,6,7,8,9\}$ aiding communications from the
source to the destination.

\begin{figure}[ht]
\centering
\includegraphics[width = 3.2in]{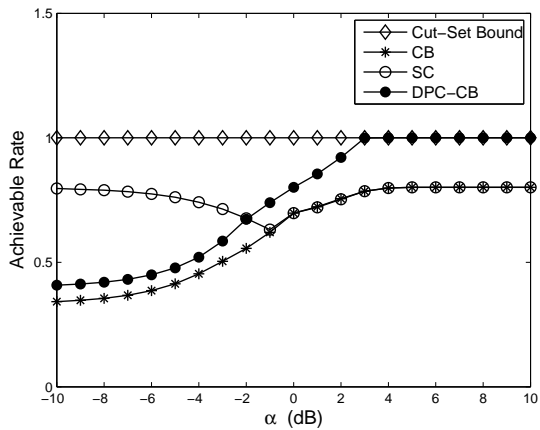}
\caption{Performance in Grid Network, $\beta = 1, \gamma = 1$, vary $\alpha$.}
\label{Fig:PerfGrid}
\end{figure}

\vspace*{-2mm}
In Fig. \ref{Fig:PerfGrid}, the gains $\beta$ and $\gamma$ are set to
1, and the gain $\alpha$ is varied. We notice that the DPC-CB scheme
approaches the capacity for a large range of values of $\alpha >
1$. The CB and SC schemes are limited by the interference at relays 4
and 5 even for large $\alpha$. For small $\alpha$, the DPC-CB and CB
schemes are limited by the common broadcast constraint at the
relays. While SC scheme can perform better, it is still limited by
interference at relays 4 and 5 compared to the cut-set bound. 

In summary, in larger networks, the choice of schedule is
important. We have used a path-based heuristic and relied on
interference-processing for approaching the cut-set bound.

\section{Conclusions}
\label{Sec:conclusions}
Based on this work, two interesting comparisons are possible for multistage half-duplex relay networks based on the cut-set bound. For the network in Fig. \ref{Fig:LNET}, the cut-set bound evaluates to $C_{pp}=\log(1+\alpha^2P/\sigma^2)$, which can be interpreted as the capacity of a point-to-point link with power constraint $P$ and channel gain $\alpha$. Using the protocols in this work, we have shown that rates up to $C_{pp}$ are achievable by multistage half-duplex relaying in the network of Fig. \ref{Fig:LNET} for certain ranges of the channel gains $\alpha$, $\beta$ and $\gamma$. A necessary condition for achieving the point-to-point capacity under the half-duplex constraint is that the source needs to be in transmit mode at all times. From our work, it appears that continuous transmission by the source and information transfer through the half-duplex relays is possible as long as there are two or more non-overlapping paths from the source to the destination (which is true in Figs. \ref{Fig:LNET} and \ref{Fig:Grid}). Further, coding in interference networks created by multiple transmitters and receivers of the relay network is crucial for enabling the information flow.

The second comparison is with full-duplex relays. The achievable rate even with full duplex relays is bounded by the sum rate across the source-broadcast cut, which is equal to $C_{pp}$, for the network in Fig. \ref{Fig:LNET}. Once again, we observe from our work that two non-overlapping paths through the relays and interference-network coding enable a half-duplex relay network to achieve the full-duplex cut-set bound for certain ranges of channel gains.


\end{document}